\documentclass[aps,amssymb,amsmath,pra, twocolumn,superscriptaddress]{revtex4}
\usepackage{amsmath}
\usepackage{graphicx}
\usepackage{amsfonts}
\usepackage{epsfig}
\usepackage{hyperref}
\usepackage{amssymb}
\usepackage{bm}
\usepackage{bbm}
\usepackage{subfigure}
\newcommand{\be}{\begin{equation}}
\newcommand{\ee}{\end{equation}}
\newcommand{\bc}{\begin{center}}
\newcommand{\ec}{\end{center}}
\newcommand{\bea}{\begin{eqnarray}}
\newcommand{\eea}{\end{eqnarray}}

\begin{document}
\title{Generic quantum walk using a coin-embedded shift operator}
\author{C. M. \surname{Chandrashekar}}
\affiliation{Institute for Quantum Computing, University of Waterloo, 
Ontario N2L 3G1, Canada}
\affiliation{Perimeter Institute for Theoretical Physics, Waterloo, ON, N2L 2Y5, Canada}

\begin{abstract}
The study of quantum walk processes has been widely divided into two standard variants, the discrete-time quantum walk (DTQW) and  the continuous-time quantum walk (CTQW).   The connection between the two variants has been established by considering the limiting value of the coin operation parameter in the DTQW, and the coin degree of freedom was shown to be unnecessary  \cite{frederick}.  But the coin degree of freedom is an additional resource which can be exploited to control the dynamics of the QW process. In this paper we present a generic quantum walk model using a quantum coin-embedded unitary shift operation $U_{C}$. The standard version of the DTQW and the CTQW can be conveniently retrieved from this generic model, retaining the features of the coin degree of freedom in both variants. 
\end{abstract}

\maketitle
\preprint{Version}

\section{Introduction}
\label{intro} 

The quantum walk (QW) as it is known today is a generalization of the classical 
random walk (CRW) developed by exploiting the aspects of quantum mechanics, such 
as superposition and interference \cite{riazanov, feynman, aharonov, meyer96, meyer2}. 
In the CRW the particle moves in the position space with a certain probability, whereas  
the QW, which involves a superposition of states, moves by exploring multiple possible paths simultaneously with the amplitudes corresponding to different  paths interfering. 
This makes the variance of the QW on a line grow quadratically with the number of steps, compared to the linear growth for the CRW.   A probabilistic result is obtained upon measurement. Several  quantum algorithms have been proposed   using   QWs~\cite{childs,
shenvi, childs1, ambainis}. Experimental implementation of the QW has been 
reported \cite{ryan, du, perets}, and various other schemes  have  
been  proposed  for  its  physical realization \cite{travaglione, rauss, 
eckert, chandra06, ma}. Beyond quantum computation, they can be used to 
demonstrate coherent quantum control over atoms, photons, or spin chains. 
The quantum phase transition using a QW is one of them  \cite{chandra07a}. Direct experimental evidence for wavelike energy transfer within photosynthetic systems has been reported, emphasizing the role of the QW \cite{ECR07}.   
\par
There are two widely studied variants of the QW,  the continuous-time quantum walk (CTQW) 
and the discrete-time quantum walk (DTQW) \cite{kempe}. In the CTQW \cite{farhi}, one  can directly define the walk on the position space, whereas in  the DTQW  \cite{andris}, it is necessary to introduce a quantum coin operation to define the direction in which the particle has to move. The results from the CTQW and the DTQW are often similar, but due to the coin degree of freedom the discrete-time variant has been shown to be more powerful than the other  in  some contexts  \cite{ambainis}, and  the coin  
parameters can be  varied to  control the dynamics of the evolution \cite{meyer96, chandra08}. To match the performance of a spatial search using the DTQW, the coin degree of freedom has been introduced in the CTQW model \cite{childs04}.  The relation between the DTQW and the CTQW remained unclear and was an open problem \cite{andris04} until the limiting value of the coin operation parameter in the DTQW was considered to establish the connection \cite{frederick}. Later, the Dirac equation-DTQW-CTQW relationship was also established \cite{frederick1}. In this construction the coin degree of freedom was shown to be unnecessary. But a coin degree of freedom is an extra resource; the parameters of the quantum coin can be exploited to control the evolution of the QW with potential applications in quantum computation \cite{chandra08} and to simulate and control the dynamics in physical systems \cite{chandra07a}. The previous closest connection between the DTQW and CTQW is the weak limit theorem for the probability density \cite{konno0205, grimmett04, gottlieb05}.
\par
The main motivation for this paper is to construct a generic QW model that will retain the features of the coin operation and establish the connection between the standard variants of the QW. Since the QW a  quantization of the classical diffusion process, it is quite natural to think in the direction of a generic model which leads to the different known variants of the QW under restrictions on the degrees of freedom of the physical system or the external resources used for implementing the QW.
\par
We construct a generic model as an extension of the DTQW model. We replace the fixed local unitary shift operator $U$ by the fixed local {\it coin-embedded shift operator} $U_{C}$. This will eliminate a separate coin toss operation on the particle to define the direction of the motion but retains the features of the coin operation. $U_{C}$ is a physically feasible construction which will reduce the generic model to the standard version of the  DTQW or the CTQW depending on the restrictions on the degrees of freedom of the initial physical system. It is well known that physical systems are not free of  environmental effects and it is shown that the QW behaviour  is  very sensitive to the environmental effect \cite{brun, kendon, chandra07}.  The environmental effects on the two operations, the coin operation $C$  and  the unitary shift operation $U$, used in the realized and most of the proposed implementable schemes of the DTQW contributes to a decrease in the decoherence time of the system. The single operation $U_{C}$ in the generic model replaces the two operations $C$  and $U$. This reduction to a single operation 
effectively contributes to an increase in the decoherence time, which in turn contributes to the increase in the number of implementable steps in the given system. The single operation $U_{C}$ also retains  the features of quantum coin parameters. 
\par
In Sec. \ref{ctdtqw} we briefly describe the standard variants of the CTQW and the DTQW.  Section \ref{genqw} discusses the construction of the generic QW model. In Secs. \ref{limitdtqw} and \ref{limitctqw} the conditions to retrieve the standard versions of the DTQW and the CTQW are presented. With a brief description of the physical implementation in Sec. \ref{imple}, we conclude in  Sec. \ref{conc}.

\section{The two variants of the QW}
\label{ctdtqw} 
We will recall both the standard variants of the quantum walk in this section. The review article by Kempe \cite{kempe} discusses them in detail. In the CTQW \cite{farhi}, the walk is defined on the {\it position}  Hilbert space $\mathcal H_{p}$ spanned  by  the  basis  state $|\psi_{x}\rangle$,  $x  \in \mathbb{Z}$. To implement the CTQW the Hamiltonian $H$ is defined such that
\be
\label{ctqw}     H|\psi_{x}\rangle     =     -|\psi_{x-1}\rangle     +
2|\psi_{x}\rangle  - |\psi_{x+1}\rangle  
\ee 
and  is made  to evolve with time $t$ by applying the transformation
\be
\label{ctqw1}
 U(t) =\exp(iHt).
 \ee
 The Hamiltonian $H$ of the process acts as the generator matrix which will transform the probability amplitude at the rate of $\gamma$ to the neighboring sites.  $\gamma$ is a fixed, time-independent, constant. 
\par
\begin{figure}
\begin{center}
\epsfig{figure=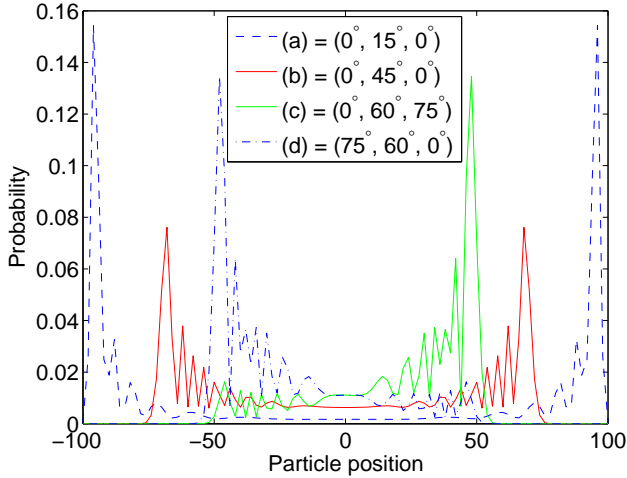, width=9.0cm}
\caption{\label{fig:qw} Distribution of the 100-step DTQW. The spread of the probability distribution for different value of $\theta$ using the operator $U_{0, \theta, 0}$, is wider for (a) = $(0,  \frac{\pi}{12},  0)$ than for (b) = $(0,  \frac{\pi}{4}, 0)$.   Biasing the walk using $\zeta$ shifts the distribution to the right, (c) = $(0, \frac{\pi}{3}, \frac{5 \pi}{12})$  and $\xi$ shifts it to the left, (d) = $(\frac{5 \pi}{12}, \frac{\pi}{3}, 0 )$. The initial state of the particle $|\Psi_{in}\rangle = \frac{1}{\sqrt 2}(|0\rangle + i |1\rangle) \otimes |\psi_{0}\rangle$.}
\end{center}
\end{figure}
The one-dimensional DTQW  \cite{andris} is defined on the Hilbert space $\mathcal  H=  \mathcal H_{c}  \otimes \mathcal H_{p}$, where $\mathcal H_{c}$ is the {\it coin} Hilbert space spanned  by the basis state of the  particle, $|0\rangle$ and  $|1\rangle$. To implement the DTQW, the 
quantum coin toss operation $C$, which in general can be an arbitrary $U(2)$ operator \cite{meyer96}, is applied on the particle at the origin in the state 
\be
\label{qw:in}
|\psi_{in}\rangle= [\cos(\delta)|0\rangle + e^{i\eta}\sin(\delta)|1\rangle]\otimes |\psi_{0}\rangle.
\ee
For the description we will consider an arbitrary three parameter $SU(2)$ operator of the form 
\be
\label{eq:coin}
 C_{\xi,\theta,\zeta}
\equiv    \left(   \begin{array}{clcr}   e^{i\xi}\cos(\theta)    &   &
e^{i\zeta}\sin(\theta)    \\     e^{-i\zeta}    \sin(\theta)    &    &
-e^{-i\xi}\cos(\theta)
\end{array} \right)
\ee
to get additional control over the evolution.
The quantum coin operation $C_{\xi, \theta, \zeta}$ is followed by  the conditional unitary shift operation
\be
\label{eq:alter} U =\exp(-2i\sigma_{z}\otimes Pl), \ee
\noindent where  $P$ being  the  momentum operator  and  $\sigma_{z}$ the 
Pauli $z$ operator corresponding to a step of length $l$. 
The eigenstates of $\sigma_{z}$ are denoted by $|0\rangle$ and $|1\rangle$. 
Therefore, $U$ in the form of the state of the particle takes the form 
\begin{eqnarray}
\label{eq:condshift}  U  =  |0\rangle  \langle 0|\otimes  \sum_{x  \in
\mathbb{Z}}|\psi_{x-1}\rangle  \langle \psi_{x} |+|1\rangle  \langle 1
|\otimes \sum_{x \in \mathbb{Z}} |\psi_{x+1}\rangle \langle \psi_{x}|.
\end{eqnarray}
The process of 
\be
W_{\xi, \theta, \zeta} =
U(C_{\xi,   \theta,  \zeta}  \otimes   {\mathbbm  1})
\ee
is iterated without resorting to an intermediate
measurement to realize a large number of steps of the QW. 
The three variable parameters of  the quantum  coin, $\xi$, $\theta$,  and $\zeta$,  can be
varied  to  change  the  probability  amplitude  distribution  in  the
position space,  Fig.(\ref{fig:qw}).  $\delta$ and  $\eta$ can be  varied to  get different initial states of the particle.   By varying the  parameter $\theta$ the variance can be
increased  or  decreased via the  functional  form
\be
\label{qwvar}
\sigma^{2}  \approx [1-\sin(\theta)]N^{2}.
\ee
For a particle with a symmetric superposition as the initial state the parameters  $\xi$ and $\zeta$ introduce asymmetry in the probability  distribution and their effect on the variance is very small. For a particle with an asymmetric superposition as the initial state, the parameters $\xi$ and $\zeta$ can be configured to obtain a symmetric probability distribution \cite{chandra08}. 

\section{ A Generic QW using the coin-embedded shift operator} 
\label{genqw} 

The generic QW model is constructed as an extension of the standard version of the DTQW. In the standard DTQW model, $\mathcal H_{c}$ is spanned by the basis state of the particle, $|0\rangle$ and $|1\rangle$, whereas for the generic model we will introduce an additional degree of freedom,  
\be
\label{coin}
\mathcal H_{c} = \mathcal H_{c_1} \otimes \mathcal H_{c_2}.
\ee
 $\mathcal H_{c_1}$ is spanned by the basis states $|0_U\rangle$ and $|1_U\rangle$  of the external resource which is used to implement a coin-embedded unitary displacement  $U_{C}$, and $\mathcal H_{c_2}$ is spanned by the basis states  of the particle. Depending on the state of the external resource and the state of the particle, the $U_{C}$ will implement the QW, eliminating the need for a separate coin toss operation after every unitary displacement. 
\par
To construct $U_{C}$, first let us consider the unitary shift operation used in the DTQW model, Eq. (\ref{eq:alter}), which can also take the form 
\begin{eqnarray}
\label{eq:alter1} U = e^{-2i\sigma_{z}\otimes  Pl} = e^{-i(|0\rangle \langle 0|
- |1\rangle  \langle 1|)\otimes  Pl} \nonumber \\  
=(|0\rangle \langle
0|\otimes e^{-iPl})(|1\rangle \langle 1|\otimes e^{iPl}).
\end{eqnarray}
\par
To embed a coin operation into the above expression, the external resource that is used to implement the unitary shift operation on the particle has to be defined such that it is local, that is, at each and every position space it is in the superposition state
\be
\label{qw:inc}
|\Psi_{U}\rangle= [\cos(\theta)|0_U\rangle + e^{i\gamma} \sin(\theta)|1_U\rangle].
\ee
If the external resource that implements the shift operator is in the
state  $|0_U\rangle$, then the particle in state $|0\rangle$ shifts to the left and the particle in state $|1\rangle$ shifts to the right.  If the external resource is in the state  $|1_U\rangle$ then the particle in state $|0\rangle$ shifts to the right and the particle in state $|1\rangle$ shifts to the left. 
\par
From the above description, the coin-embedded shift operation $U_{C}$ takes the form

\begin{widetext}
\begin{eqnarray}
\label{eq:alter1a} 
 U_{C} =  \left \{ |0_{U}\rangle  \langle   0_{U}|\otimes
\exp \left [-i(|0\rangle \langle 0|
- |1\rangle  \langle 1|)\otimes  Pl \right ] \right \} 
\times \left \{  |1_{U}\rangle \langle  1_{U}|\otimes \exp \left [ i(|0\rangle \langle 0|
- |1\rangle  \langle 1|)\otimes  Pl \right ] \right \}.
\end{eqnarray}
Therefore, $U_{C}$ in the form of the state of the external resource and the particle can be written as
\begin{eqnarray}
\label{eq:alter1c} 
 U_{C} =   |0_{U}\rangle  \langle   0_{U}|\otimes \left ( |0\rangle  \langle 0|\otimes  \sum_{x  \in
\mathbb{Z}}|\psi_{x-1}\rangle  \langle \psi_{x} |+|1\rangle  \langle 1
|\otimes \sum_{x \in \mathbb{Z}} |\psi_{x+1}\rangle \langle \psi_{x}| \right ) \nonumber \\
+ |1_{U}\rangle  \langle   1_{U}|\otimes \left ( |0\rangle  \langle 0|\otimes  \sum_{x  \in
\mathbb{Z}}|\psi_{x+1}\rangle  \langle \psi_{x} |+|1\rangle  \langle 1
|\otimes \sum_{x \in \mathbb{Z}} |\psi_{x-1}\rangle \langle \psi_{x}| \right )
\end{eqnarray}
The operation $U_{C}$ on the initial state of the system is of the form
\begin{eqnarray}
\label{initial1} 
\begin{array}{ll}
|\Psi_{in} \rangle &= |\Psi_{U} \rangle \otimes |\psi_{p}\rangle \otimes |\psi_{0}\rangle 
 = [\cos(\theta)|0_U\rangle + e^{i\gamma} \sin(\theta)|1_U\rangle]  
  \otimes  [\cos(\delta)|0\rangle + e^{i\eta}\sin(\delta)|1\rangle]\otimes |\psi_{0}\rangle
\end{array}
\end{eqnarray}
\end{widetext}

\noindent and implements the first step of the generic QW; here $|\psi_{p}\rangle \otimes |\psi_{0}\rangle$ is the state of the particle at the origin (position). Hereafter, we will write the state of the particle position after $t$ steps as $|\Psi_{t}\rangle$.  Since $|\Psi_{U}\rangle$ is a local state of the external resource,
after implementing $U_{C}$, the state of the particle position unentangles from the external resource to again entangle with the resource state $|\Psi_{U}\rangle$ in the new position. This can be written as
\be
({\mathbbm  1} \otimes |\Psi_{t}\rangle) =  U_{C} \left (  |\Psi_{U}\rangle  \otimes  {\mathbbm  1} \right )\left ( {\mathbbm  1} \otimes |\Psi_{t-1}\rangle \right ).
\ee
Therefore, irrespective of the internal state of the particle, $U_{C}$ moves the particle in the superposition of the position space. Note that the role of the coin operation is completely retained in the above construction through the external resource which implements $U_{C}$. By choosing an equal superposition state of the external resource, Eq. (\ref{qw:inc}), the distribution of the Hadamard walk can be retrieved. 
\par
If unit time is required to implement each step then to implement $t$ steps $U_{C}(t) = U_{C}^{t}$.  Therefore,  the  wave function after time $t$  can be written as
\be 
({\mathbbm  1} \otimes |\Psi_{t}\rangle) = [ U_{C} ( |\Psi_{U}\rangle  \otimes   {\mathbbm  1} ) ]^{t} ({\mathbbm  1} \otimes  |\Psi_{0}\rangle).  
\ee 
The probability of the particle being in position $x$ is 
\be
P_{x}(t)= |\langle x | \psi_{x,t}\rangle|^2.
\ee
  By choosing a different linear combination of the initial state of the particle and the external resource implementing $U_{C}$, the probability distribution in the position space can be controlled as is done using separate coin operations in the DTQW model. 

\section{Retrieving the standard versions from the generic model}
\label{limitqw}
\subsection{DTQW}
\label{limitdtqw}
If the external resource implementing $U_{C}$ is not in a superposition, that is,  if it is in one of its basis states $|0_{U}\rangle$ or $|1_{U}\rangle$, then the Hilbert space, Eq. (\ref{coin}), $\mathcal H_{c} \equiv \mathcal H_{c_1}$.
Therefore $U_{C}$, Eq. (\ref{eq:alter1a}),  reduces to
\be 
\label{nunit1}
U_{C}= \left ( |0\rangle \langle 0|\otimes e^{ \mp i  Pl } \right)
 \left ( |1\rangle \langle 1|\otimes
e^{ \pm i  P l } \right ),
\ee
a unitary sift operator of the standard version of the DTQW, Eq. (\ref{eq:condshift}).  Therefore, by introducing the quantum coin operation $C_{\xi, \theta, \zeta}$, the standard version of the DTQW and all its properties can be recovered.      
\par

\subsection{CTQW}
\label{limitctqw}

If the initial state of the particle is not in a superposition, that is, if it is in one of its basis states $|0\rangle$ or $|1\rangle$, then the Hilbert space, Eq. (\ref{coin}), $\mathcal H_{c} \equiv \mathcal H_{c_2}$.
Therefore $U_{C}$, Eq. (\ref{eq:alter1a}),  reduces to
\be 
\label{nunitCTQW}
U_{C}= \left ( |0_{U}\rangle \langle 0_{U}|\otimes e^{ \mp i  Pl } \right)
 \left ( |1_{U}\rangle \langle 1_{U}|\otimes
e^{ \pm i  P l } \right )
\ee
\begin{eqnarray}
\label{nunitCTQWa} U_{C} = \exp \left [\mp i(|0_{U}\rangle \langle 0_{U}|
- |1_{U}\rangle  \langle 1_{U}|)\otimes  Pl \right ].
\end{eqnarray}

If it takes unit time for each $U_{C}$ operation then after time $t$
$U_{C}(t)$ can be written as
\be
 U_{C}^{t}  = \exp \left [ \mp i(|0_{U}\rangle \langle 0_{U}|
- |1_{U}\rangle  \langle 1_{U}|)\otimes  Plt \right ].
\label{nunitCTQWb}
\ee
Since $|0_{U}\rangle$ and $|1_{U}\rangle$ are the states of the external resource used to displace the particle, the above expression reveals the effect of the state of the external resource on the particle, 
\begin{eqnarray}
\label{nunitCTQWc}
\left [(|0_{U}\rangle \langle 0_{U}|- |1_{U}\rangle  \langle 1_{U}|)\otimes Pl \right ](|\Psi_{U}\rangle \otimes \psi_{0}\rangle ) \nonumber \\ 
 =\alpha |0_{U}\rangle \otimes |\psi_{-1} \rangle +  \beta|1_{U}\rangle \otimes |\psi_{+1}\rangle,
\end{eqnarray}
where $\alpha$ and $\beta$ are the coefficients of the states $|0_{U}\rangle$ and $|1_{U}\rangle$ and $|\psi_{0}\rangle$  is the state of the particle at position $0$. Since the external resource is a local state, Eq. (\ref{nunitCTQWc}) is
\be
\equiv  \alpha |\psi_{-1}\rangle + \beta |\psi_{+1}\rangle
\ee

Therefore, Eq. (\ref{nunitCTQWb}) can be written in the form

\be
U_{C}^{t}  \equiv \exp(\pm iH_{L}t),
\label{nunitCTQWd}
\ee
where $H_{L}$ is the local Hamiltonian in the position space.
The probability amplitude transition rate $\gamma$ is related to the state of the external resource. By choosing an arbitrary superposition state of the external resource, Eq. (\ref{qw:inc}), different transition rates $\gamma_{1}$ for the left and $\gamma_{2}$  for the right can be obtained; this is also equivalent to the introduction of the coin degree of freedom to the standard CTQW model \cite{childs04}. Thus the generic quantum walk model can be reduced to the standard version of the CTQW. The CTQW happens irrespective of the state of the particle. All the features of the coin operation in the standard version of the DTQW can be retrieved in this version of the CTQW.  This construction also makes the connection between the DTQW and the CTQW very straightforward. \\
\vskip 0.2in
\section{Physical implementation}
\label{imple}

A simple physical system can be considered in which the polarized light can act as a coin-embedded unitary shift operator $U_{C}$. It can be conditioned such that the vertically polarized light ($|0_{U}\rangle$) will shift the particle in state $|0\rangle$ to the left and the particle in state $|1\rangle$ to the right. Horizontally polarized light  ($|1_{U}\rangle$) shifts the particle in state $|1\rangle$ to the left and the particle in state $|0\rangle$ to the right. Therefore light in a coherent superposition of the vertical and horizontal polarizations $[\cos (\theta) |0_{U}\rangle + e^{i \gamma}\sin (\theta) |1_{U}\rangle]$, can implement the $U_{C}$ on a particle. The other physical advantage of using the generic model for the implementation of the QW is the possibility of increasing the decoherence time by reducing the number of operations needed to implement each step of the QW.  The environmental effects on the two operations, the coin operation $C$ and the displacement operation $U$ in the physical system contribute to a reduction in the decoherence time. Replacement of the two operations $C$ and $U$ by a single operation contributes to a decrease in the environmental effects on the system and increases its decoherence time.    
\section{Conclusion} \label{conc}

In summary, we have constructed a generic QW model by embedding the coin operation into the unitary shift operator $U_{C}$. The generic model retains the features of the coin operation and establishes the connection between the two standard versions of the DTQW and the CTQW.  When the external resource that implements $U_{C}$ is not in a superposition of its basis states, the standard version of the DTQW can be retrieved by introducing an addition coin operation.  When the particle on which the generic model is implemented is not in a superposition of its internal states, the CTQW is retrieved along with the features of the coin degree of freedom. This makes the CTQW reproduce all the features of the standard version of the DTQW. This model, along with establishing a link between the two versions of the QW, can also play a prominent role in increasing the decoherence time of the system and hence increasing the realizable number of steps in a given physical system. 
\\
\bc
\bf{Acknowledgement}
\ec

I thank Raymond Laflamme for his encouragement and support, and Andrew Childs and Sarvagya Upadhyay for comments on the manuscript. I also acknowledge  Mike and Ophelia Lezaridis for  financial support at IQC.


\end{document}